\documentclass[a4paper,12pt]{article}
\usepackage{makeidx}
\usepackage{amsmath}
\usepackage{latexsym}
\usepackage{amssymb}

\usepackage{graphicx}
\usepackage{float}
\begin{document}

\thispagestyle{empty}
\vskip 12pt
\hbox to 16.5 true cm{\hfill TOKAI-HEP/TH-0333\ \ \ \ \ \ \ }
\vskip 32pt

\begin{center}
{\Large\bf Dilatonic Inflation and SUSY Breaking\\ in String-inspired  Supergravity}
\\
\vspace{16pt}
Mitsuo J. Hayashi${}^*$, Tomoki Watanabe${}^\dagger$, Ichiro Aizawa and Koichi Aketo
\\
\vspace{16pt}
{\it\footnotesize Department of Physics, Tokai University, 1117, Kitakaname, Hiratsuka, 259-1292, Japan}

{\footnotesize ${}^*$mhayashi@keyaki.cc.u-tokai.ac.jp;
${}^\dagger$2aspd004@keyaki.cc.u-tokai.ac.jp}
\\

\vspace{24pt}
{\bf Abstract}
\end{center}
The theory of inflation will be investigated as well as supersymmetry breaking in the context of supergravity, incorporating the target-space duality and the nonperturbative gaugino condensation in the hidden sector. 
We found an inflationary trajectory of a dilaton field and a condensate field which breaks supersymmetry at once. 
The model satisfies the slow-roll condition which solves the $\eta$-problem. When the particle rolls down along the minimized trajectory of the potential $V(S,Y)$ at a duality invariant point of $T=1$, we can obtain the $e$-fold value $\sim 57$. And then the cosmological parameters obtained from our model well match the recent WMAP data combined with other experiments.
This observation suggests one to consider the string-inspired supergravity as a fundamental theory of the evolution of the universe as well as the particle theory. 

\newpage
\addtocounter{page}{-1}

 The theory of inflation are still the most promising theory of the early universe before the big bang\cite{ref:1}.
Remarkably, on February 11, 2003, NASA's Wilkinson Microwave Anisotropy Probe (WMAP) was able to capture the cosmic microwave background in the infant universe, which reveals the afterglow of the big bang. ``The data show the first generation of stars shines in the universe first ignited only 200 million years after the big bang and pegs the age of the universe at 13.7 billion years. The WMAP team found that the big bang and inflation theories continue to be true. The contents of the universe include four percent of ordinary matter, 23 percent of an unknown type of dark matter, and 73 percent of a mysterious dark energy which acts as a sort of anti-gravity to realize the accelerating expansion of the universe"\cite{ref:2}. In theoretical point of views, people had introduced the quintessence for naming the origin of the dark energy\cite{ref:3}, and the quintessence inflation has been investigated\cite{ref:4}. Recently, the authors in Ref.\cite{ref:4}  pointed out that the underlying particle theory, which will be responsible for the present quintessential behavior of our universe, is likely to have contact with supersymmetry, supergravity or superstring theory.
 
 As far as the 4D, $N=1$ supergravity\cite{ref:5} can play an elementary role in the theory of  the spacetime and the particles\cite{ref:6}, it can also be essential in the theory of the early universe as an effective field theory. However, supergravity has been confronting with the difficulties, such as the $\eta$-problem and the supersymmetry breaking mechanism. Various trials to prevail over these difficulties has been well studied by many authors\cite{ref:7}. 
 In this paper we will revisit the no-scale supergravity as the low energy effective theory of strings\cite{ref:8,ref:9}. Hidden sector supergravity breakdown appears as the most promising 4D, $N=1$ supergravity setup for the low energy effective theory as that of beyond the standard model\cite{ref:10}.
 
 Let us, in this paper, investigate the possibility that both the hidden sector supersymmetry breakdown and inflation occur at once.
There had been investigations on gaugino condensation and supersymmetry breaking, in the cases of pure super Yang-Mills \cite{ref:11,ref:12,ref:14} or that with massive hidden sector matter\cite{ref:13}.     
We will concentrate on the former case by revisiting the paper by Ferrara {\it et al.} of Ref.\cite{ref:12}.
We found that the interplay between condensate gauge-singlet scalar $Y$ and the dilaton field $S$ can give rise to sufficient inflation by $S$ and supergravity breakdown by condensate field $Y$. 
\newline

First of all, for the self-containedness, we will review the construction of the effective theory of gaugino condensation, incorporating the target-space duality, following to Ref.\cite{ref:12}, where the gaugino condensation has been described by a duality-invariant effective action for the gauge-singlet gaugino bound states coupled to the fundamental fields like the dilaton $S$ and moduli $T$. On the other hand, in Ref.\cite{ref:11}, the gaugino-condensate has been replaced by its vacuum expectation value to yield a duality-invariant ``truncated" action that depends on the fundamental fields only.  The equivalence between these two approaches had been proved in Ref.\cite{ref:13}.

 Assuming that the compactification of the superstring theory preserves $N=1$ supersymmetry, the effective theory should be of the general type of $N=1$ supergravity coupled to gauge and matter fields\cite{ref:5}. The most general theory of this type is described by the Lagrangian:
\begin{equation}
\mathcal{L}=-\frac{1}{2}[e^{-K/3}S_0\bar{S}_0]_D +[S_0^3W]_F +[f_{ab}W^a_\alpha \epsilon^{\alpha\beta}W^b_\beta]_F,
\end{equation}
where $K$ is the K\"{a}hler potential, $W$ is the superpotential and $f_{ab}$ is the gauge kinetic function. All these functions depend on the fundamental chiral matter fields with conformal weight 0, while the gauge superfield $W^a_\alpha$ has conformal weight 3/2. $S_0$ is the chiral compensator superfield of conformal weight 1, which ensures the correct weights for each of the three terms. All physical quantities depend on $K(\phi,\phi^\ast)$ and $W(\phi)$ through the only relevant combination:
\begin{equation}
\mathcal{G}=K+\ln|W|^2.
\end{equation}

The tree-level K{\"a}hler potential is of the no-scale type and is given by:
\begin{equation}
K=-\ln \left(S+S^\ast\right)
-3\ln \left(T+T^\ast-|\Phi_i|^2\right),
\end{equation}
where $\Phi_i$ represent chiral matter fields. The gauge function is $f_{ab}=\delta_{ab} S$ \cite{ref:9}.

The target space modular group $PSL(2,{\bf Z})$ acts
on the complex scalar $T$ as:
\begin{equation}
T\to\frac{aT-ib}{icT+d},\quad ad-bc=1,
\end{equation}
whereas $S$ remains invariant at the tree-level and $\Phi_i\to(icT+d)^{-1}\Phi_i$.

According to the underlying conformal field theory, all interactions are 
target space modular invariant at any order of string perturbation theory, 
so that the effective string action should be modular invariant as well.
Demanding modular invariance of $\mathcal{G}$ to guarantee modular invariance
of the theory, the superpotential must transform as modular weight $-3$ under
$PSL(2,{\bf Z})$:

\begin{equation}
W\to\frac{W}{(icT+d)^3},
\end{equation}
up to a $T$-independent phase factor.

By introducing the composite superfield:
\begin{equation}
U=\delta_{ab}W^a_\alpha \epsilon^{\alpha\beta}W^b_\beta,
\end{equation}
we may construct the effective theory of gaugino condensation following 
Ref.\cite{ref:14}. As $U$ has conformal weight 3, we will introduce the superfield:
\begin{equation}
\tilde{U}=U/S_0^3,
\end{equation}
of conformal weight 0. Then, in terms of $\tilde{U}$, the gauge kinetic term of Eq.(1) gives the contribution to the tree-level superpotential:
\begin{equation}
\tilde{W}^{\rm tree}=S\tilde{U}.
\end{equation}

The one-loop modification of the superpotential should be $S$ independent and of the form:
\begin{equation}
\tilde{W}^{\rm one-loop}=\tilde{U}f(\tilde{U},T),
\end{equation}
with $f$ a modular function of weight 0, while $\tilde{U}$ has a modular form of weight $-3$. Following the general arguments based on the fulfilment of anomalous Ward identities\cite{ref:14}, $f$ is required to contain the term:
\begin{equation}
\frac{\beta_0}{96\pi^2}\log{\tilde{U}},
\end{equation}
where $\beta_0$ is the coefficient of $\beta$-function.
The logarithm of a composite field had been originally introduced to construct the supermultiplet structure which can reproduce the chiral and scale anomaly as analogy from QCD, though the $F$-term does not get vacuum expectation value.
It is well known that, under the assumption that the target space modular group is an exact symmetry of string theory, strong constraints appear in the form of nonperturbative four-dimensional effective supergravity action providing a link to modular function theory.
In fact, so as to satisfy the modular invariance of $f$, a modular form of weight 3 has to be multiplied to $\tilde{U}$ under the argument of this logarithm. In order to avoid unphysical singularities, the argument of the logarithm is required to have neither zeroes nor poles in the upper half-plane of complex $iT$. This demand leaves us with the unique form:
\begin{equation}
\eta(T)^6=e^{-\pi T/2} \prod^{\infty}_{n=1}(1-e^{-2\pi nT})^6,
\end{equation}
where $\eta(T)$ is the Dedekind $\eta$-function, defined by:
\begin{equation}
\eta(T)=e^{-2\pi T/24} \prod^{\infty}_{n=1}(1-e^{-2\pi nT}).
\end{equation}
$\eta(T)$ has the modular transformation property of weight 1/2 under
$PSL(2,{\bf Z})$:
\begin{equation}
\eta(T)\to(icT+d)^{1/2}\eta(T).
\end{equation}

It is convienient to introduce the chiral superfield $Y$ defined by $\tilde{U}=Y^3$, which has modular weight $-1$.
Then, the effective no-scale type K{\"a}hler potential and the effective superpotential that incorporate modular invariant one-loop corrections are given by:
\begin{equation}
K=-\ln \left(S+S^\ast\right)
-3\ln \left(T+T^\ast-|Y|^2-|\Phi_i|^2\right),
\end{equation}
\begin{equation}
W=3bY^3\ln\left[c\>e^{S/3b}\>Y\eta^2(T)\right],
\end{equation}
where $b=\frac{\beta_0}{96\pi^2}$, and we have disregarded the correction to the dilaton field $S$ in the first term.
The effective gauge kinetic function is given by:
\begin{equation}
f=S+3b\ln[Y\eta^2(T)]+{\rm const}.
\end{equation}
Since $\langle S+S^\ast\rangle =\alpha^\prime m_{\rm P}^2$, the choice:
\begin{equation}
[e^{-K/3}S_0\bar{S}_0]_{\theta=\bar\theta=0}=[S+\bar{S}]_{\theta=\bar\theta=0}
\end{equation}
corresponds to the conventional normalization of the gravitational action:
\begin{equation}
\mathcal{L}_{\rm grav} \sim [e^{-K/3}S_0\bar{S}_0]_{\theta=\bar\theta=0}R.
\end{equation}
Now, the scalar potential is in order:
\begin{eqnarray}
V(S,T,Y)&=&\frac{3(S+S^\ast)|Y|^4}{(T+T^\ast-|Y|^2)^2}
\Bigg(3b^2 \left|1+3\ln\left[c\>e^{S/3b}\>Y\eta^2(T)\right]\right|^2
\nonumber\\
&&{}+\frac{|Y|^2}{T+T^\ast-|Y|^2}
\left|S+S^\ast-3b\ln\left[c\>e^{S/3b}\>Y\eta^2(T)\right]\right|^2
\nonumber\\
&&{}+6b^2|Y|^2\left[2(T+T^\ast)\left|\frac{\eta^\prime(T)}{\eta(T)}\right|^2
+\frac{\eta^\prime(T)}{\eta(T)}
+\left(\frac{\eta^\prime(T)}{\eta(T)}\right)^\ast\right]\Bigg).
\end{eqnarray}
Here, all the matter fields are set to zero for simplicity.
The potential is explicitly modular invariant and can be shown to be stationary at the self-dual points $T=1$ and $T=e^{i\pi /6}$\cite{ref:15}.
\newline

Now we will show an inflationary trajectory of a dilaton field and a condensate field which breaks supersymmetry satisfying the slow-roll condition and can solve the $\eta$-problem. For the sake of simplicity, we will set $S=S^\ast$ and $Y=Y^\ast$.
First, the plot of the potential $V(S,Y)$ at a self-dual fixed point $T=1$ is shown at Fig. 1; we can see a valley of the potential and a stable minimum of $V_Y(S)=0$ at $(Y_{\rm min},S_{\rm min})\sim (0.00646, 0.435)$. 
\begin{figure}[H]
\begin{center}
\includegraphics[scale=.8]{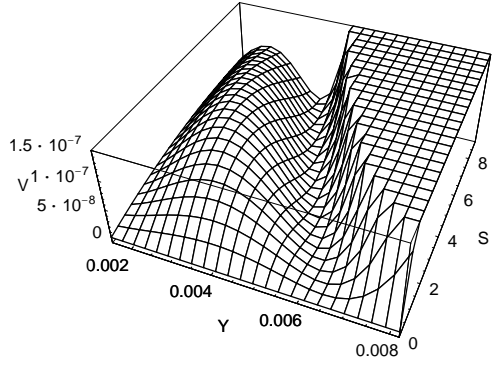}
\end{center}
\caption{The plot of $V(S,Y)$ at fixed $T=1$ (self-dual point) with $c=183,\ b=5.5$.}
\end{figure}
Therefore, the particle should roll down along a trajectory minimized with respect to $Y$, and then 
we can conclude that inflation arises by the evolution of dilaton field $S$, provided the slow-roll condition is satisfied.

The inflationary trajectory will be well approximated by the equation:
\begin{equation}
Y_{\rm min}(S)\sim 0.00663e^{-S/16.2},
\end{equation}
which is slightly different from Ferrara {\it et al.}'s approximated equation\cite{ref:12}: $|Y|=(ce^{1/3}|\eta (T)|^2)^{-1}e^{-{\rm Re}S/3b}$. 
In Fig. 2, we have shown a plot of $V(S)$ minimized with respect to $Y$.
As shown by Ferrara {\it et al.}\cite{ref:12}, supersymmmetry is broken by the hidden sector gaugino condensation because 
$\langle |F| \rangle \propto \langle |\lambda\lambda |\rangle\neq 0$. 
\begin{figure}[H]
\begin{center}
\includegraphics[scale=.8]{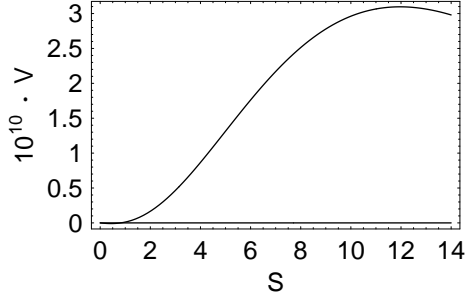}
\end{center}
\caption{The plot of $V(S)$ minimized with respect to $Y$. The minimum value of the potential is $V(S_{\rm min})\sim -9.3\times 10^{-13}$.}
\end{figure}
Our main purpose of this paper is to prove that the dilaton field plays the role of inflaton field. The slow-roll parameters (in Planck units $m_{\rm P}/\sqrt{8\pi}=1$) are defined by:
\begin{equation}
\epsilon_\alpha=\frac{1}{2}\left(\frac{\partial_\alpha V}{V}\right)^2
\quad ,\quad 
\eta_{\alpha\beta}=\frac{\partial_\alpha\partial_\beta V}{V}.
\end{equation}
The slow-roll condition demands both values to be lower than 1. We understand that it is the end of inflation, when one of the slow-roll parameters reach the value 1. After passing through the minimum of the potential, ``matters" may be produced with the critical density, i.e. $\Omega=1$.
The values of $\epsilon_S$ and $\eta_{SS}$ are obtained numerically in Figs. 3 and 4; we find the condition is well satisfied, and the $\eta$-problem can just be avoided.
\begin{figure}[H]
\begin{center}
\begin{minipage}{.45\linewidth}
\includegraphics[scale=.8]{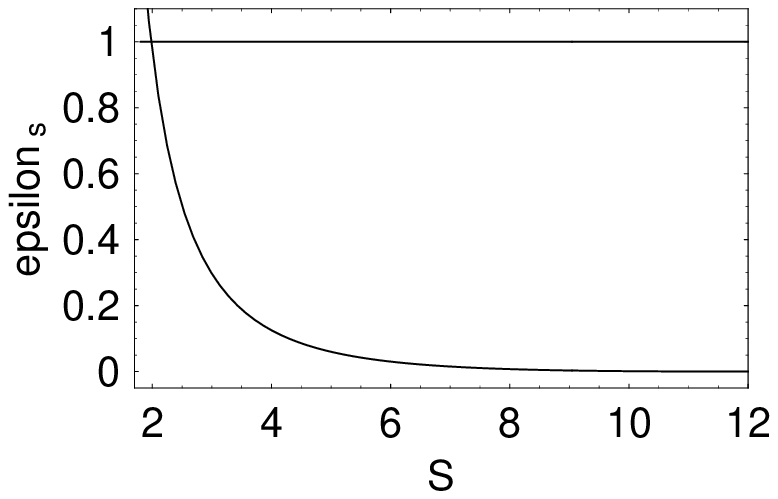}
\caption{The plot of $\epsilon_S$.}
\end{minipage}
\begin{minipage}{.45\linewidth}
\includegraphics[scale=.8]{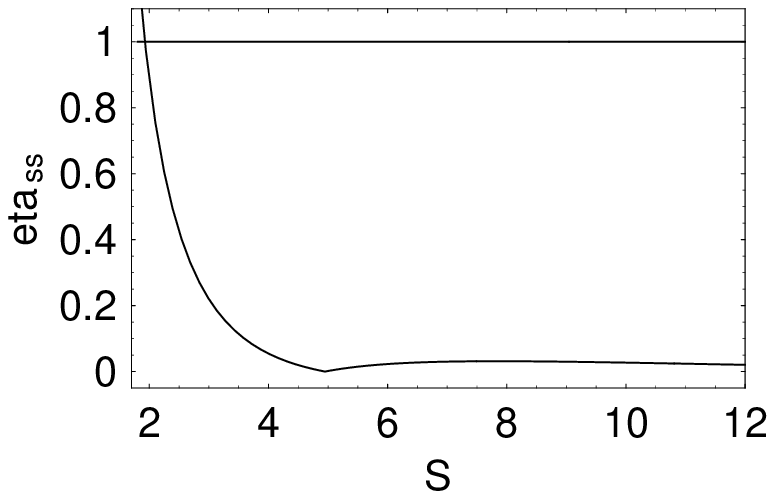}
\caption{The plot of $|\eta_{SS}|$.}
\end{minipage}
\end{center}
\end{figure}

Let us now examine whether our scenario is viable or not in comparison with the recent measurements\cite{ref:2}.

 Number of $e$-folds at which a comoving scale $k$ crosses the Hubble scale $aH$ during inflation is given by\cite{ref:1}:
\begin{equation}
N(k)\sim 62-\ln\frac{k}{a_0H_0}-\frac{1}{4}\ln\frac{(10^{16}\>{\rm GeV})^4}{V_k}+\frac{1}{4}\ln\frac{V_k}{V_{\rm end}},
\end{equation}
where we assume $V_{\rm end}=\rho_{\rm reh}$. We focus the scale $k_*=0.05\> {\rm Mpc^{-1}}$ and the inflationary energy scale is $V\sim10^{-10}\sim(10^{16}\>{\rm GeV})^4$ as shown in Fig. 2, therefore the number of $e$-folds which corresponds to our scale must be around 57. On the other hand, using the slow-roll approximation, $N$ is also calculated by:
\begin{equation}
N\sim -\int^{S_2}_{S_1}\frac{V}{\partial V}dS.
\end{equation}
We could have obtained the number $\sim 57$, by integrating from $S_{\rm end}\sim 1.98$ to $S_*\sim 10.46$, fixing the parameters $c=183\ {\rm and}\ b=5.5$, i.e. our potential has the ability to produce the cosmologically plausible number of $e$-folds. Here $S_*$ is the value corresponding to $k_*$.

Next, a scalar spectral index standing for a scale dependence of the spectrum of density perturbation and its running are defined by:
\begin{equation}
n_s-1=\frac{d\ln \mathcal{P_R}}{d\ln k}\ ,\ \alpha_s=\frac{dn_s}{d\ln k}.
\end{equation}
These are approximated in the slow-roll paradigm as:
\begin{equation}
n_s(S)\sim 1-6\epsilon_S+2\eta_{SS}\ ,\  \alpha_s(S)\sim 16\epsilon_S\eta_{SS} -24\epsilon_S^2-2\xi^2_{(3)},
\end{equation}
where $\xi_{(3)}$ is an extra slow-roll parameter that includes trivial third derivative of the potential. Substituting $S_*$ into these, we have $n_{s*}\sim0.95$ and $\alpha_{s*}\sim -4\times10^{-4}$. Because $n_s$ is not equal to 1 and $\alpha_{s}$ is negligible, our model supports the model with tilted power law spectrum. The value of $n_{s*}$ is consistent with the recent observations; the best fitting of them (WMAPext, 2dFGRS and Lyman $\alpha$) for power law $\Lambda$CDM model suggests $n_s(k_*)=0.96\pm0.02$\cite{ref:2}.

Finally, estimating the spectrum of the density perturbation\cite{ref:16} caused by slow-rolling dilaton:
\begin{equation}
\mathcal{P_R}\sim\frac{1}{12\pi^2}\frac{V^3}{\partial V^2},
\end{equation}
we find $\mathcal{P_R}_*\sim2.1\times10^{-9}$. This result matches the measurements as well\cite{ref:2,ref:19}. Incidentally speaking, the energy scale $V\sim10^{-10}$ is also within the constrained range\cite{ref:19}.
  
 Now we conclude that inflation and supersymmetry breaking occur at once by the interplay between the dilaton field as inflaton and the condensate gauge-singlet field.
 
 For further investigations, (a) We should consider on the effects of the hidden sector massive matter over inflation and the supersymmetry breaking\cite{ref:13}. (b) it will be interesting to understand what kind of phenomena are obtained from the S-duality invariant theory\cite{ref:6}. Furthermore, (c) the dark matter and the dark energy problem appears to be the most important and exciting as well as the production of ordinary matter\cite{ref:2,ref:3,ref:4}. (d) gravitino, inflatino and axion production and their effects should be traced\cite{ref:17}. (e) brane world cosmology might be promising\cite{ref:18}, (f) $M$-theoretical approach seems also important\cite{ref:6,ref:10}. These problems will be our further tasks.

\eject
\end{document}